\documentclass[aps,pra,twocolumn,amsmath,nofootinbib]{revtex4-1}
\usepackage{epsfig}
\usepackage[colorlinks,linkcolor=blue,anchorcolor=blue,
citecolor=blue,urlcolor=blue,
breaklinks=true]{hyperref}
\usepackage{graphics}
\usepackage{color}
\usepackage{graphicx}% Include figure files
\usepackage{dcolumn}% Align table columns on decimal %point
\usepackage{bm}% Bold math
\usepackage{cases}
\usepackage{appendix}
\usepackage{ulem}

\begin{document}
\author{Jin-Xiang Xue$^{1}$}
\author{Chuan-Xun Du$^{1,3}$}
\email{Email: duchuanxun@lyu.edu.cn}
\author{Chengchao Liu$^{1}$}
\author{Liu Yang$^{1}$}
\author{Yong-Long Wang$^{1,2,3}$}
\email{Email: wangyonglong@lyu.edu.cn}
\address{$^{1}$ School of Physics and Electronic Engineering, Linyi University, Linyi, 276005, China}
\address{$^{2}$ Department of Physics, Nanjing University, Nanjing 210093, China}
\address{$^{3}$ National Laboratory of Solid State Microstructures, Department of Materials Science and Engineering, Nanjing University, Nanjing, 210093, China}
\title{The Non-reciprocity of Multi-mode Optical Directional Amplifier Realized by Non-Hermitian Resonator Arrays}

\begin{abstract}
In the present paper, a multi-frequency optical non-reciprocal transmission is first realized by using a non-Hermitian multi-mode resonator array.We find that the non-reciprocity can be used to route optical signals, to prevent the reverse flow of noise, and find that the multi-frequency can be used to enhance information processing. In terms of the Scully-Lamb model and gain saturation effect, we accomplish a dual-frequency non-reciprocal transmission by introducing nonlinearity into a linear array of four-mode resonators. For example, a directional cyclic amplifier is constructed with non-reciprocal units. As potential applications, the non-reciprocity optical systems can be employed in dual-frequency control, parallel information processing, photonic integrated circuits, optical devices and so on. 
\bigskip
		
\noindent PACS Numbers: XXXXXXX
\end{abstract}
\maketitle
	
\section{Introduction}\label{1}
Optical systems can be widely applied in scientific research and engineering, such as optical communication, optical imaging, optical sensing and so on. Particularly, the photons are employed in information calculation and transmission for the high bandwidth and low loss. \cite{58,77}. Recently, the non-Hermitian optical system is of great interests of experts in optics\cite{58,14,22,59}, because that the gain and loss are easily controlled in optical systems.  Specifically, a circulator and directional amplifier of fiber-coupled silicon dioxide micro-resonators  was realized\cite{771} for a single frequency optical signal.  While multi-frequency directional cyclic amplifiers still need further investigations in optical nonlinearity.
	
The reciprocity of an optical system means that  the response of the system does not change  when the light wave is backpropagated along the direction of propagation\cite{61}. The reciprocity is valuable in many optical systems and devices, but in special cases the non-reciprocity plays an important role in optical devices, in which light waves propagate considerably different forward and backward\cite{6,20,63}. The non-reciprocity is important in photonic information processing to route optical signals or prevent the reverse flow of noise. The non-reciprocal characteristics of light can be applied to optical devices\cite{74(2),79,75},  in which signal transmission and reverse isolation can be achieved in a specific direction, and the interference effect can be effectively reduced to protect the signal source from noise interference. As a result, the non-reciprocal optical devices are more accurate and reliable,  which is of great significance in the research of communication technology, information processing, precision measurement and topological photonics\cite{79,73,74(1),76,81,75}. Various non-reciprocal devices were realized for single-frequency\cite{771,49,52,82}, but the multi-channel information processing capability still needs to improve the non-reciprocal optical devices that are easily integrated in optical chips. Therefore, the micro-non-magnetic devices need further investigations to own multi-frequency non-reciprocal characteristics\cite{64}.
	
In the presence of magnetic field, the non-reciprocity can be generated by a variety of low power non-magnetization methods\cite{32,33,34,35,36}, which can be achieved by magnetic field\cite{40,42,64}, the dynamic modulation of dielectric constant\cite{34,40,37,38,75}, optical nonlinearity\cite{81,43,44,45} and so on. In terms of dynamic modulation permittivity and optical nonlinearity, the non-reciprocal mechanics was applied to superconducting microwave circuits\cite{49,46,47,48,50,51}. For potential applications, it should be first considered that is integration, stability, controllability and other aspects. As the magnetic field absences, the previous achievements become more difficult. However, the high integration, strong stability and easy controllability can be easily achieved in optical devices by using the optical nonlinear effect\cite{76}. For a parity time-symmetric photonic device consisting of two resonators\cite{737}, the light propagation was analyzed by the non-Lindbladian master equation of the Scully-Lamb laser model. By semi-classical approximation, two nonlinear coupled differential equations were obtained for the field evolution in a cavity, which can explain the non-reciprocity\cite{497}, and demonstrate that the gain saturation effect is the key mechanism of non-reciprocity\cite{737}. It is worthwhile to notice that the known non-reciprocity phenomenon mostly appears at the non-divergent peak of a single frequency. In our work\cite{83}, a series of PT-symmetric one-dimensional multimode resonator arrays were built and the transmission spectra was discussed. In the system, the non-reciprocal phenomenon can be achieved by the gain saturation effect. We tried to replace the transition point of the PT unbroken phase by that of broken phase in the parameter space, and realized multiple non-divergent peaks in the transmission spectrum. The results may provide a way to design multi-frequency non-reciprocal devices.
	
In the present paper, we will try to construct a dual-frequency non-reciprocal optical system by using a non-Hermitian multimode resonator array. In terms of the Scully-Lamb model, we will give a nonlinear equation by introducing a nonlinear term into the original linear equation.  With the gain saturation effect\cite{737} and the non-Hermitic resonator array model, the known two-mode resonator system can be extended to a four-mode resonator system. And then the optical non-reciprocity of two frequencies can be achieved. This phenomenon can not only be applied to photodiodes, but also the dual-frequency non-reciprocity phenomenon and the parallel processing of multiple information. And the system can be used to build multi-mode optical directional amplifier and isolators\cite{6,32,51,54,55,57}. The amplifier structure, shown in Figure 1, consists of an optical fiber coupled with a silicon dioxide ring microresonator with four ports. The signal can be input from one port, the output signal can be detected from the others, and the signal will be amplified with different times for different output ports. If only two of the ports are taken, the function of the isolator can be realized. The function of directional amplifier can be realized by controlling the frequency mismatch between the input light and the cavity resonance. Compared with other directional amplifiers, this system has several advantages, such as dual-frequency control, circulability, and easy integration.
	
The paper is organized as follows. In Sec. II, the basic model and the linear dynamics equations are briefly introduced. In Sec. III, the theoretical basis of the system is given, and the dual-frequency non-reciprocal transmission can be achieved by a four-mode resonator system, due to the presence of the nonlinear gain saturation. In Sec. IV, a non-reciprocal double-frequency system is constructed and investigated.  In Sec. V, conclusions and discussions are given.

\section{Basic Model and Linear Dynamic Equation}
As a basic model, a ring resonator array will be considered in this section, and its linear dynamic equations will be discussed. With the linear equations, the position and type of resonance peaks can be determined. In terms of the Scully-Lamb model, by introducing nonlinear terms, the considered model can give rise to non-reciprocal transmission phenomena. The result plays an important role in the present paper. Specifically, the model introduced nonlinear terms can be taken as fundamental elements to construct complex multi-mode non-reciprocal transmission systems and subsequent multi-mode unidirectional amplifiers.
	
\begin{figure}[htbp]
\centering
\includegraphics[width=0.4\textwidth]{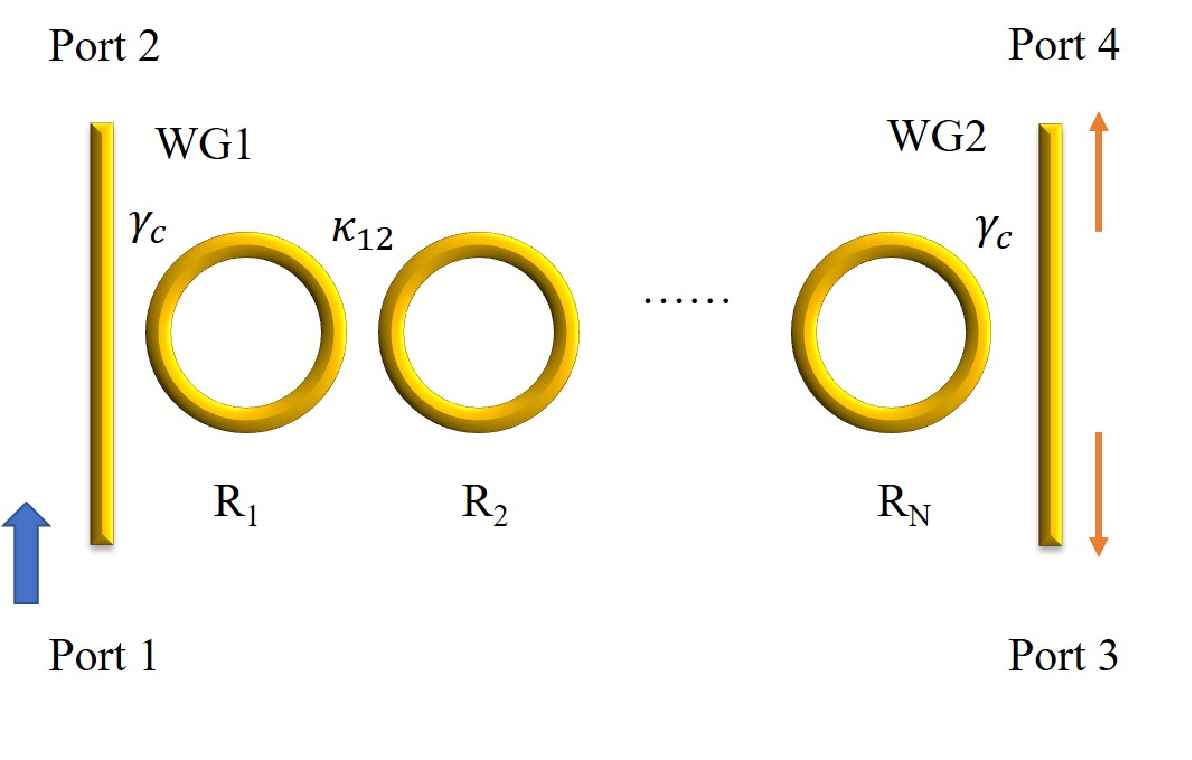}
\caption{\footnotesize Setup of the one-dimensional array of ring resonators. The resonators labeled as $R_1 \sim R_N$ are coupled with each other with coupling strength $\kappa_{ij}$. The coupling loss between waveguide WG1(WG2) and resonator $R_1(R_N)$ is $\gamma_c$. }\label{fig2}
\end{figure}	
	
The considered system is a linear array of $N$ coupled resonators depicted in Fig. \ref{fig2}. In comparing with the work referenced in \cite{497}, it is significant that moves from a two-mode to a multi-mode configuration. The passive resonators, which are lossy, can be crafted from silica, while the active resonators, which provide gain, can be made from an erbium-ion-doped silica film. This film is fabricated by the sol-gel technique, where the solution is prepared by blending tetraethoxysilane with isopropanol alcohol, water, and hydrochloric acid. The resonators are in turn denoted as $R_1 \sim R_N$, and they have the same frequency $\omega_0$, and their couplings just occur between adjacent resonators, the coupling parameter can be described by the real-valued parameter $\kappa_{i,j}$. The internal loss of the $i$th resonator is given by $\gamma'_i=\omega_0/Q_i$ being $Q_i$ the Q-factor, and the gain provided by optical pumping is denoted as $g'_i$. In addition, the first and last resonators experience a loss $\gamma_c$ due to their connection with the waveguides WG1 and WG2, respectively. The net gain or loss for the $i$th resonator can be encapsulated in a single parameter  $\gamma_i=g'_i-\gamma'_i-\gamma_c(\delta_{i,1}+\delta_{i,N})$ , where gain is indicated by $\gamma_i>0$ and loss by $\gamma_i<0$. The probe signal is introduced through port 1, and the resulting signal can be monitored at port 3 or 4, contingent upon the value of $N$ being even or odd. The coupling strength between the first resonator and the input signal's driving field, characterized by power $P$ and frequency $\omega_{\text{in}}$, is represented by  $\epsilon$, with the relationship defined as $\epsilon\equiv\sqrt{\gamma_cP/\hbar\omega_{\text{in}}}$.
	
When the input power $P$ does not surpass the threshold for gain saturation effects \cite{741}, and as long as the resonator peaks remain bounded, the gain within the active cavity behaves linearly. As a consequence, the dynamics of the optical field can be expressed as a set of linear equations \cite{497,498,737},
	
\begin{widetext}		
\begin{equation}\label{aDyEq}
\begin{aligned}
\frac{\partial a_{n}}{\partial t}= \left  \{
\begin{aligned}
&-i\omega_0 a_{1}+ \gamma_1 a_{1}+i\kappa_{1,2} a_{2}-\epsilon {\rm exp} (-i\omega_{\text{in}} t) &(n=& 1),\\
&-i\omega_0 a_{n}+ \gamma_n a_{n}+i\kappa_{n,n+1} a_{n+1}+i\kappa_{n,n-1}a_{n-1} &(1<n<& N),\\
&-i\omega_0 a_{N}+ \gamma_N a_{N}+i\kappa_{N,N-1} a_{N-1} &(n=& N),
\end{aligned}
\right.
\end{aligned}
\end{equation}		
By substituting $a_n=A_n{\rm exp} (-i\omega_{\text{in}} t)$ into Eq.\eqref{aDyEq},  the dynamical equations for the field amplitude can be obtained,
\begin{equation}\label{ADyEq}
\begin{aligned}
\frac{\partial A_{n}}{\partial t}= \left  \{
\begin{aligned}
& i\Delta_{\text{in}} A_{1}+ \gamma_1 A_{1}+i\kappa_{1,2} A_{2}-\epsilon &(n= & 1),\\
& i\Delta_{\text{in}}A_{n}+ \gamma_n A_{n}+i\kappa_{n,n+1} A_{n+1}+i\kappa_{n,n-1}A_{n-1} &(1<n<& N),\\
& i\Delta_{\text{in}}A_{N}+ \gamma_N A_{N}+i\kappa_{N,N-1} A_{N-1} &(n=& N),
\end{aligned}
\right.
\end{aligned}
\end{equation}		
where $\Delta_{\text{in}}=\omega_{\text{in}}-\omega_0$ is the frequency-detuning parameter. In the steady state ${\partial A_{n}}/{\partial t}=0$, from Eq.\eqref{ADyEq} one can obtain a series of linear equations in the following form 
\begin{equation}\label{AnLEq}
\begin{aligned}
& i\Delta_{\text{in}} A_{1}+ \gamma_1 A_{1}+i\kappa_{1,2} A_{2}=\epsilon &(n=1),\\
& i\Delta_{\text{in}}A_{n}+ \gamma_n A_{n}+i\kappa_{n,n+1} A_{n+1}+i\kappa_{n,n-1}A_{n-1}=0 &(1<n<N),\\
& i\Delta_{\text{in}}A_{N}+ \gamma_N A_{N}+i\kappa_{N,N-1} A_{N-1}=0 &(n=N).
\end{aligned}
\end{equation}
\end{widetext}
It is apparent that $A_N$ can be completely determined by solving Eqs.\eqref{AnLEq}. 
In the semiclassical approximation, the coupling constant $\epsilon$ can be expressed via the amplitude $A_{\text{in}}$ of the input driving field as $\epsilon\equiv\sqrt{\gamma_c}A_{\text{in}}$.~\cite{737} Then the transmission spectrum can be calculated by
	
\begin{equation}\label{Trans}
T(\Delta_{\text{in}})=\left|\frac{A_{\text{out}}}{A_{\text{in}}} \right|^2,
\end{equation}
where $A_{\text{out}}=\sqrt{\gamma_c}A_N$ and $A_{\text{in}}=\epsilon/\sqrt{\gamma_c}$.
It is easy to check that when the parameter setting of $\gamma_i$ and $\kappa_{i,j}$ changes, the distribution of resonant peaks of the graph of $T(\Delta_{\text{in}})$ will become different. To study how non-reciprocity happens on these peaks, one can rewrite Eqs.\eqref{ADyEq} and \eqref{AnLEq} in matrix form. As a subsequence, the question of finding resonant peaks is converted into an eigenvalue problem about the coefficient matrix.	
	
\begin{figure}[htbp]
\centering
\includegraphics[width=0.4\textwidth]{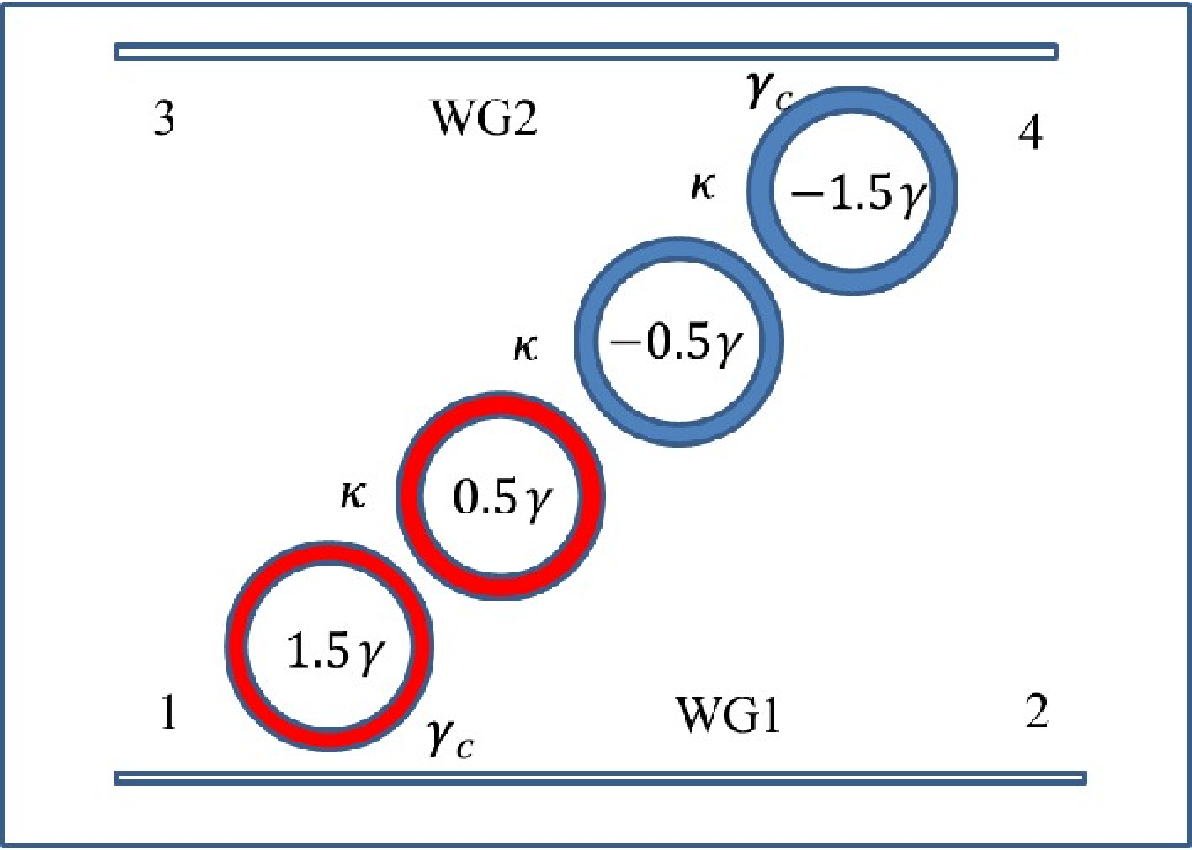}
\caption{\footnotesize A four-mode resonator array with balanced gain and loss. The resonators are coupled adjacent to each other with coupling strength $\kappa$. The coupling loss between waveguide WG1(WG2) and resonator $R_1(R_N)$ is $\gamma_c$.}\label{fig3}
\end{figure}	
	
By defining a vector $\alpha=(A_1,A_2,...,A_N)^T$ with elements being the field amplitudes, and by introducing $\beta=(\epsilon,0,...,0)^T$  describing the input condition, we can rewrite Eq.\eqref{ADyEq} in the form $i{\partial A_{n}}/{\partial t}=\hat H\alpha-\beta$, where the coefficient matrix $\hat{H}$ reads	
\begin{equation}\label{H}
\hat H=\begin{pmatrix}
\Delta_{\text{in}}+i \gamma_{1} & \kappa_{1,2} & \cdots & 0 \\
\kappa_{2,1} & \Delta_{\text{in}}+i \gamma_{2} & & \vdots \\
\vdots & & \ddots & \\
0 & \cdots & & \Delta_{\text{in}}+i \gamma_{N}
\end{pmatrix},
\end{equation}
which is known as the Hamiltonian or evolution matrix~\cite{497,498,737}. With $\alpha$, $\beta$ and $\hat{H}$, Eq.\eqref{AnLEq} can be simplified as $\hat H \alpha=\beta$, and then the steady-state solution of $\alpha$ can be given by $\alpha=\hat H^{-1}\beta$. It is worthwhile to notice that the solution of $\alpha$ is protected by the non-singularity of $\hat{H}$, $|\hat H|\ne 0$. If $\hat{H}$ is singular, the vector $\alpha$ and the transmission coefficient $T(\Delta_{in})$ could not be determined. Furthermore, the amplitude $A_N$ can not be determined. This divergence in $T(\Delta_{in})$ indicates that the system cannot achieve a steady state, which leads to an infinite amplification of the light field during its transmission between resonators, akin to a lasing process. This phenomenon is referred to as a resonance state in the continuum (RIC) ~\cite{765}, and is also known as a spectral singularity ~\cite{766,767}. For a particular set of $\gamma_i$ and $\kappa_{i,j}$, the frequencies are indicative of the divergent peaks in the transmission spectrum, which is termed supermodes. For the supermodes of the photon frequencies might be complex due to solutions involving imaginary parts. In what follows, for the sake of reality the series of non-divergent peaks will correspond to the real parts of these supermodes. Thus, the issue of resonant peaks is reduced to an eigenvalue problem. This methodology is analogous to the Siegert boundary condition approach, which is commonly used to identify resonant states in open quantum systems~\cite{766,764,872,873}, where resonant states align with the eigenstates of the Schr\"{o}dinger equation under certain boundary conditions~\cite{764}. This correspondence is not unexpected given the formal equivalence between the Schr\"{o}dinger equation and the classical wave equation.
	
Practically, the light intensity cannot be infinitely amplified. Actually, the divergent peaks should be replaced by a series of sharp but finite peaks. Intriguingly, non-reciprocal transmission phenomena emerge at non-divergent peaks~\cite{497}. The discovery inspires physicists to investigate the underlying mechanisms. As a considered plausible theory,  the gain saturation theory was given by Ievgen I. Arkhipov's group \cite{737}. In the theory, the nonlinear equations is derived  from the Scully-Lamb laser model. It is worth noticing that the nonlinear equations combine the linear model’s simplicity and the nonlinear model’s accuracy.
	
%Continuing from the previous section, we now turn our attention to the transformation of the linear equations into a nonlinear form. This conversion is based on the principles outlined in reference [737], which builds upon the Scully-Lamb model. According to this literature, the presence of gain in a resonant cavity necessarily introduces corresponding nonlinear terms. Given this theoretical foundation, we can extend the linear equations derived in the first section to incorporate these nonlinear effects. This expansion not only accounts for the more complex behavior observed in real-world systems but also provides a more accurate representation of the underlying physics. By including these additional terms, we aim to capture the subtleties and nuances that arise due to the system's nonlinearity, ultimately leading to a more comprehensive understanding of its dynamics, and the nonlinear equations can be written as:
\section{Gain Saturate Effect and Multi-mode Non-reciprocal Transmission}
	
%从文献737得知，Scully-Lamb模型对线性方程的修正，可以整理为每个增益项后面加一个非线性项。
%而观察双模PT系统的数值结果（见附录），可以得到两个结论，1、非线性方程能够预测出非互易峰，2、非互易峰出现在PT对称破缺区域的非发散峰。
%再根据我们组的前期工作的一个结论，即非发散峰的位置与复平面上的超模在实轴上的投影位置相一致，进而我们建立了如下模型。
	
According to Ref.\cite{737}, the Scully-Lamb model's modification to linear equations can be  accomplished by adding a nonlinear term into each gain term. Upon examining the numerical outcomes of the dual-mode PT system (see Appendix), two key insights emerge: one is the nonlinear equations predicting non-reciprocal peaks, the other is the non-reciprocal peaks  manifesting within the non-divergent peaks region of PT symmetry breaking. As the conclusion obtained in Ref.\cite{83}, the location of non-divergent peaks aligns with the projection of supermodes onto the real axis in the complex plane. The model is constructed as in Fig.\ref{fig3}.
%该模型由四个一维排列的谐振腔组成，彼此相等的耦合系数以及对称的增益损耗系数保持了它的PT 对称性。可以看出它是图2所示模型的一个特例。因而它的线性速率方程可以表示为Eq.(2) 的一个特殊形式，即：
	
The model consists of four ring resonators in a one-dimensional configuration, with equal coupling coefficients and symmetric gain-loss coefficients. The particular constructions preserve  the PT symmetry. As a special case of the model depicted in Fig.\ref{fig2}, from Eq.\eqref{ADyEq} the specific linear rate equations can be given by 
\begin{equation}\label{SADyEq}
\begin{aligned}
& \frac{d}{d t} A_1=i \Delta A_1+1.5 \gamma A_1-i \kappa A_2-\varepsilon; \\
& \frac{d}{d t} A_2=i \Delta A_2+0.5 \gamma A_2-i \kappa A_1-i \kappa A_3 ; \\
& \frac{d}{d t} A_3=i \Delta A_3-0.5 \gamma A_3-i \kappa A_2-i \kappa A_4 ; \\
& \frac{d}{d t} A_4=i \Delta A_4-1.5 \gamma A_4-i \kappa A_3,
\end{aligned}
\end{equation}
and the corresponding coefficient matrix is in the following form
\begin{widetext}
\begin{equation}\label{H4}
H_4=\left(\begin{array}{cccc}
\Delta_{\text {in }}+i 1.5 \gamma & \kappa & 0 & 0 \\
\kappa & \Delta_{\text {in }}+i 0.5 \gamma & \kappa & 0 \\
0 & \kappa & \Delta_{\text {in }}-i 0.5 \gamma & \kappa \\
0 & 0 & \kappa & \Delta_{\text {in }}-i 1.5 \gamma.
\end{array}\right).
\end{equation}
	
\begin{figure}[htbp]
\centering
\includegraphics[width=1\textwidth]{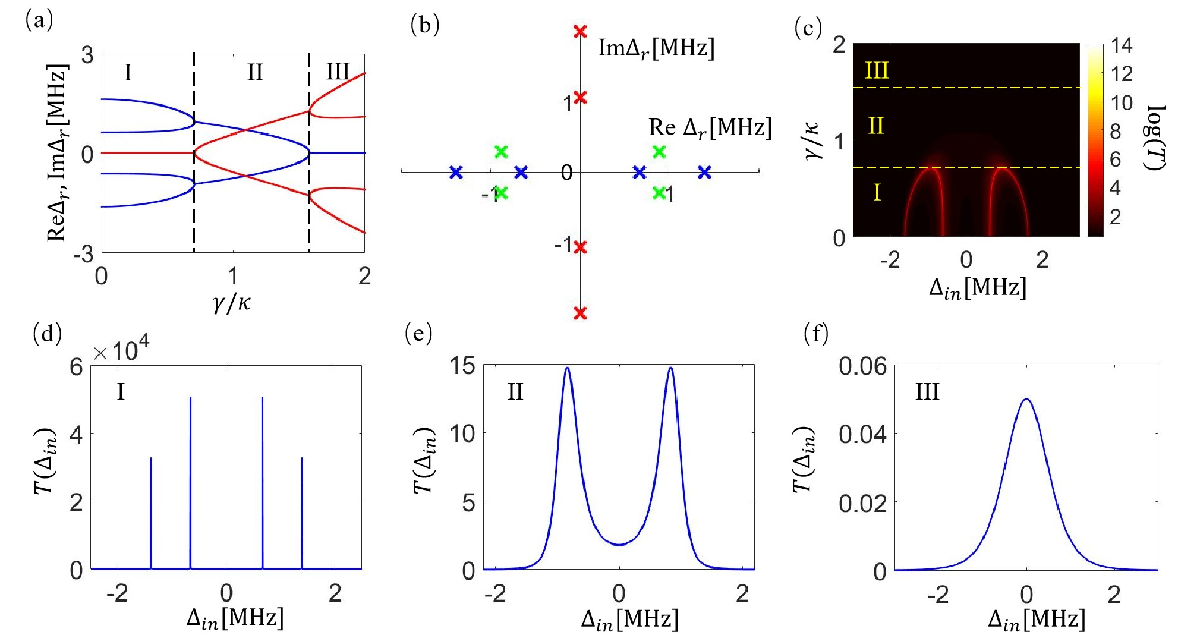}
\caption{\footnotesize (a) The real (blue) and imaginary (red) parts of supermodes of $\hat H_4$ as  functions of $\gamma/\kappa$. (b)The distribution of supermode $\Delta_r$ on the complex plane with $\kappa=1$MH at $\gamma=0.5$MHz (blue), $\gamma=0.8$MHz (green), $\gamma=1.8$MHz (red). (c)The variation of transmission spectrum as a function of $\gamma/\kappa$ and $\Delta_{\text{in}}$, where the logarithm of $T(\omega_l)$ is shown as the brightness. (d)(e)(f) Transmission spectrum against $\Delta_{\text{in}}$ in the  region I ((d),$\gamma=0.5$MHz), region II ((e),$\gamma=0.8$MHz) and region III ((f),$\gamma=1.8$MHz). Wherein $\kappa=\gamma_c=1$MHz,$\epsilon=0.1$MHz.}\label{fig4}
\end{figure}
\end{widetext}
	
By numerically solving the secular equation, the supermodes $\Delta_r$ of $H_4$ are obtained, whose real parts (blue) and imaginary parts (red) of $\Delta_r$ are described in Fig.\ref{fig4}(a). $\gamma$ is small, the supermodes $\Delta_r$ are purely real which corresponds to the blue points in Fig.\ref{fig4}(b). In region III, $\gamma$ is large, $\Delta_r$ are purely imaginary, which corresponds to the red ones. While in region II, the real parts and imaginary parts of $\Delta_r$ are all non-zero, they are denoted by the green points in Fig.\ref{fig4}(b). In other words, the parameter space is divided into three regions: in I, $\mathcal{PT}$ unbroken $\Delta_r$ as real; II being transition region $\Delta_r$ as complex; and in III, $\mathcal{PT}$ broken $\Delta_r$ as imaginary.
	
In different regions the transmission spectrum has different behaviors as shown in Figs.\ref{fig4}(c)-(f). In region I, the number of divergent peaks equals to that of the supermodes as in Fig.\ref{fig4}(d), and in region III, there is a single non-divergent peak on $\Delta_{\text{in}}=0$ as in Fig.\ref{fig4}(f). However, in region II, there are more than one non-divergent peaks in the spectrum as in Fig.\ref{fig4}(e). Comparing with the green points in Fig.\ref{fig4}(b), we find that the locations of non-divergent peaks coincide with the projections of supermodes on to the real-axis. In terms of previous speculations, the multiple non-divergent peaks may appear in the region II successfully.
	
\begin{widetext}
		
\begin{figure}[htbp]
\centering
\includegraphics[width=0.9\textwidth]{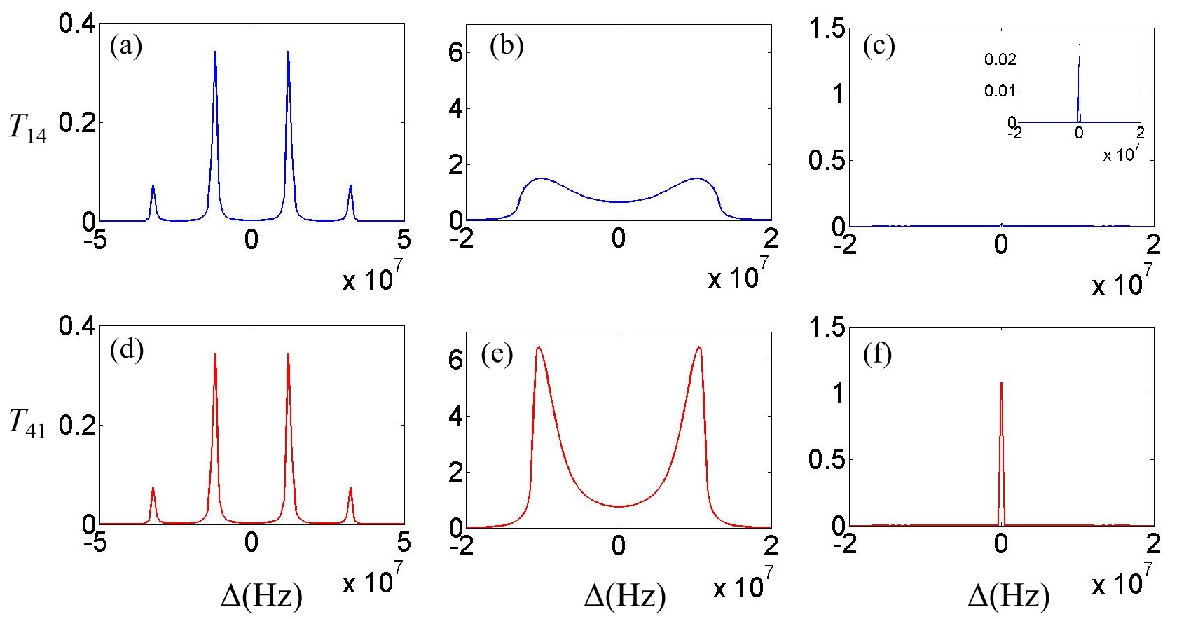}
\caption{Transmission spectrum $T_{14}$ and $T_{14}$ against $\Delta_{\text{in}}$ with $\epsilon=2GH, \gamma=1MH, B=0.01Hz$, (a) (d) at $\kappa=20MH$,(b) (e) at $\kappa=1.1MH$,(c) (f)at $\kappa=0.1MH$.}\label{fig5}
\end{figure}
\end{widetext}

When the gain saturation effect is taken into account, the linear equation will be reformulated into the following nonlinear form,
\begin{equation}\label{ANDyEq}
\begin{aligned}
& \frac{d}{d t} A_1=i \Delta A_1+1.5 \gamma A_1-i \kappa A_2-\varepsilon-B\left|A_1\right|^2 A_1 ; \\
& \frac{d}{d t} A_2=i \Delta A_2+0.5 \gamma A_2-i \kappa A_1-i \kappa A_3-B\left|A_2\right|^2 A_2 ; \\
& \frac{d}{d t} A_3=i \Delta A_3-0.5 \gamma A_3-i \kappa A_2-i \kappa A_4 ; \\
& \frac{d}{d t} A_4=i \Delta A_4-1.5 \gamma A_4-i \kappa A_3 ;
\end{aligned}
\end{equation}
According to Eq.\eqref{H4}, the transmission spectrum can be easily solved with $\varepsilon=2GH, \gamma=1MH$ and $B=0.01Hz$ at $\kappa=20MH, 1.1MH, 0.1MH$, the results are depicted in Fig.\ref{fig5}. The results confirm two predictions: One is that the divergent peak from the linear equation becomes a sharp but finite non-divergent peak without non-reciprocal phenomenon, as shown in Figs.\ref{fig5}(a)(d), which corresponds to the region I in Fig.\ref{fig4}(a). The other is that the  divergent peak becomes a non-divergent one but with non-reciprocal transmission phenomenon, as shown in Figs.\ref{fig5}(b)(e), which corresponds to the region II in Fig.\ref{fig4}(a).
	
The emergence of multi-frequency non-reciprocal phenomena as in Figs.\ref{fig5}(b) and (e)that is the most important result in the present paper. In the next section, we will further discuss a Dual-Frequency Circulator by using the nonlinearized model.
	
\section{Construction of a Dual-Frequency Circulator}
In this section, we will introduce a nonlinear term into the original linear equation, and will extend a two-mode cavity system to a four-mode cavity system, in which a two-frequency non-reciprocity phenomenon can be achieved. Interestingly, the non-reciprocity of the nonlinear model can be employed to construct an optical directional amplifier providing simultaneously multi-mode optical channels with different magnifications. In other words, the non-reciprocity can provide an access to manipulate simultaneously multi-mode optical channels, which could be applied in optical chip integration.
	
\begin{figure}[htbp]
\centering
\includegraphics[width=0.4\textwidth]{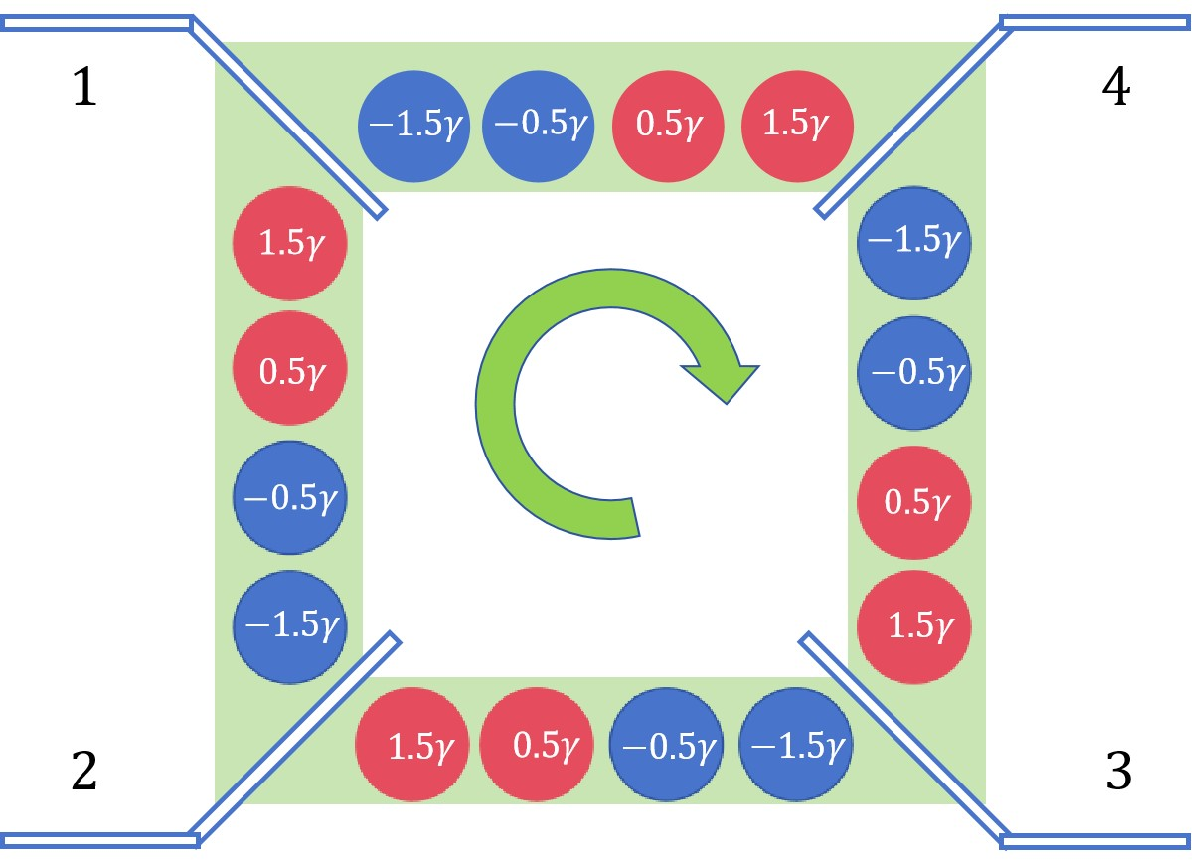}
\caption{\footnotesize Schematic diagram of multi-mode optical directional amplifier (the blue cavity is a loss cavity and the red cavity is a gain cavity). The device has four ports, of which any two ports can be used as signal input and output.}\label{fig6}
\end{figure}
	
\begin{figure}[htbp]
\centering
\includegraphics[width=0.3\textwidth]{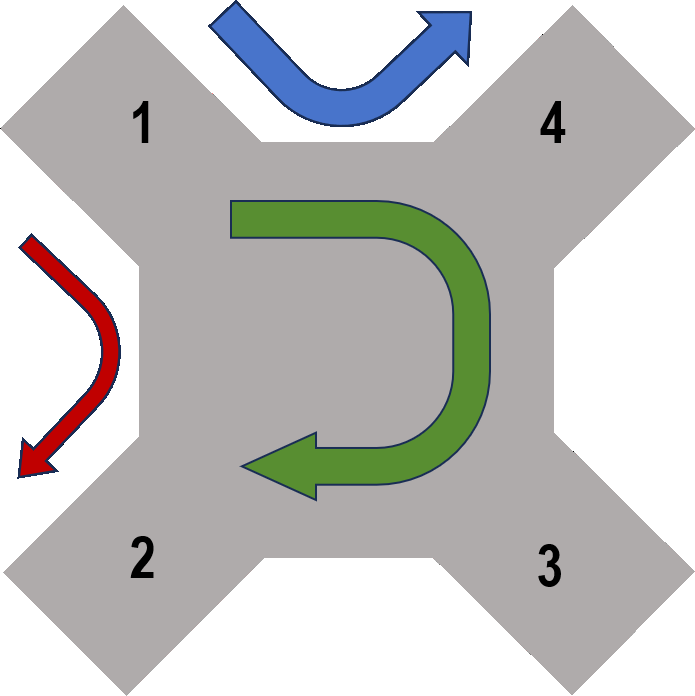}
\caption{\footnotesize Schematic diagram of optical nonreciprocal directional amplifier. When the signal light is routed in a clockwise direction, as shown by the blue arrow, there is a large amplification factor between ports. When the signal light is routed in a counterclockwise direction, as shown by the red arrow, there is a small amplification factor between ports. The green arrow represents our ideal signal routing direction.}\label{fig1}
\end{figure}
	
As an example, a double-mode non-reciprocal optical directional amplifier can be constructed by using four basic units as described in Fig.\ref{fig6}. Each unit consists of four ring echo wall resonators as in Fig.\ref{fig3}. The non-reciprocal amplifier has four ports, and any two of them can be used as signal input and output respectively. When the signal enters the optical fiber from one of the ports, it will be propagated in two routes (clockwise and counterclockwise) at the same time as sketched in Fig.\ref{fig1}. Specifically, in the amplifier any port can be taken as an output to detect signal, and can be taken as an input to continue the signal execution into the next unit. The particular construction can simultaneously provide many channels for signal transmission.
	
\begin{figure}[htbp]
\centering
\includegraphics[width=0.5\textwidth]{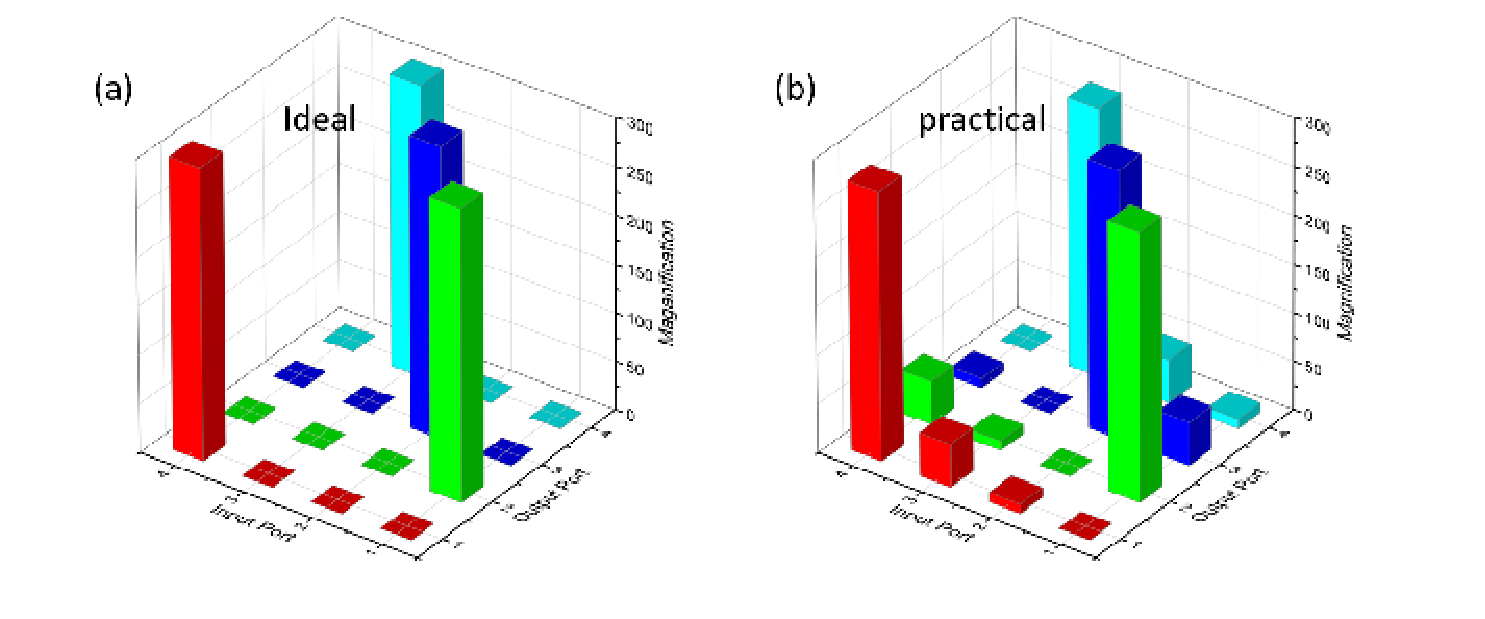}
\caption{\footnotesize Amplification coefficient matrix of multi-mode optical directional amplifier for single cycle. (a) The ideal magnification matrix. $M_{12}=M_{23}=M_{34}=M_{41}=300$ (b) The actual resulting magnification matrix. When the signal light is incident from any port, it is transmitted in a clockwise direction, and the light intensity is continuously amplified. $M_{11}=M_{22}=M_{33}=M_{44}=1$ }\label{fig7}
\end{figure}
	
In the example, it is obvious non-reciprocal that the transmission spectrum of the signal propagates in clockwise and in anticlockwise. For the sake of simplicity, we assume that the optical signal propagates a single cycle, and we introduce a metric $M$ to describe the transmission between the ports for the double-mode non-reciprocal amplifier. The element of metric $M_{ij} (i,j=1,2,3,4)$ describes the transmission between the $i$th port and the $j$th port, where the $i$th port is input and the $j$th port is output. Specifically, we can describe the transmission for the port 1 to the port 2 as $M_{12}=T_{41}^3+T_{14}$, that for the port 1 to the port 3 as $M_{13}=T_{41}^2+T_{14}^2$, for that of the port 1 to the port 4 as $M_{14}=T_{41}+T_{14}^3$, where $T_{41}$ and $T_{14}$ are the transmission of the basic unit as shown in Figs.\ref{fig5}(b)and(e). By numerical calculation, we can obtain that $M_{12}$ is much larger than $M_{14}$, that is the output of port 2 being much larger than that of port 4. This result directly demonstrates the existence of non-reciprocity in the example. More significant, the numerical results demonstrate that the optical signal is amplified 300 times through a single cycle as described in Fig.\ref{fig7}. The results may provide a way that the non-reciprocal amplifier may be collaborated in complex opto-mechanical structures and implemented in integrated photonic circuits.

\section{Conclusion}
In summary, based on the Scully-Lamb model and the photomechanical interaction, we have constructed a non-reciprocal unit that is a system coupled by four resonators and two optical fibers. Particularly, it is non-reciprocal that in the constructed system the transmission rate has different values in a particular direction and its opposite. For the presence of many ports and simultaneously distinguishable transmission rates in the non-reciprocal unit, the dual-frequency non-reciprocity phenomenon can realize parallel transmission and processing of information, and can be applied to photonic integrated circuits or optical devices\cite{49,32,46,47}. As communication applications, one can construct optical isolator, circulators and directional amplifier\cite{32,51,54,55,57}, which can be used to control optical signals in a particular direction at specific frequencies. As magneto-optical storage, one can design magneto-optical materials and devices\cite{56} with the non-reciprocal units.

For an example, the multi-mode optical directional amplifier with four ports was built by using the basic units. For a particular input port, the rest three output ports can have distinguishable amplified times of optical signals, which results from the non-reciprocal. These results may provide an approach for the more and more complex requirements of signal in communication and optical integrated circuits. Moreover, the example can be arbitrarily extended to more ports that can be integrated on chip with other circuit components in barrier-free combination.

\section*{Acknowledgments}
This work is jointly supported by the National Nature Science Foundation of China (Grants, No. 12475019, No.1247030364), and National Lab of Solid State Microstructure of Nanjing University (Grants No.M35040, No.M35053 and No.M37014).

\section*{APPENDIX}
Gain saturate effect is a pivotal phenomenon in laser physics where the intensity
of the intracavity field becomes significant enough to reduce the net gain of the
active medium. As the Scully-Lamb quantum laser model is applied to parity-time symmetric whispering-gallery microcavities, the gain saturation effect plays an important role in explaining the non-reciprocity in light propagation. When the input driving field intensity increases, the gain ability of medium will diminish due to saturation, which equivalently decrease the amplification of the signal field. This nonlinear mechanism results in a direction-dependent transmission behavior, which is essential for the non-reciprocal light transmission in the broken PT-symmetric phase. The model demonstrates that gain saturation modifies the eigenmodes and exceptional points of the system, by introducing an additional loss mechanism that depends on the field intensity and propagation direction. In Ref.\cite{737}, the authors employ the Scully-Lamb quantum laser model to derive a set of rate equations that describe the dynamics of light propagation in parity-time symmetric whispering-gallery microcavities. Starting from a non-Lindbladian master equation that accounts for gain, loss, and gain saturation, they apply the semiclassical approximation to transform quantum operators into classical amplitudes. By calculating the time derivatives of these amplitudes and incorporating the effects of gain saturation, at a set of nonlinear coupled differential equations are obtained, which can elucidate the interplay between the intracavity fields and the response of photonic system, and reveal the nonreciprocal transmission of light. This theoretical framework bridges the quantum and classical descriptions, and provides a way to learn the action of gain saturation on the non-reciprocal behavior.
\begin{widetext}
	
	\begin{figure}[htbp]
		\centering
		\includegraphics[width=1\textwidth]{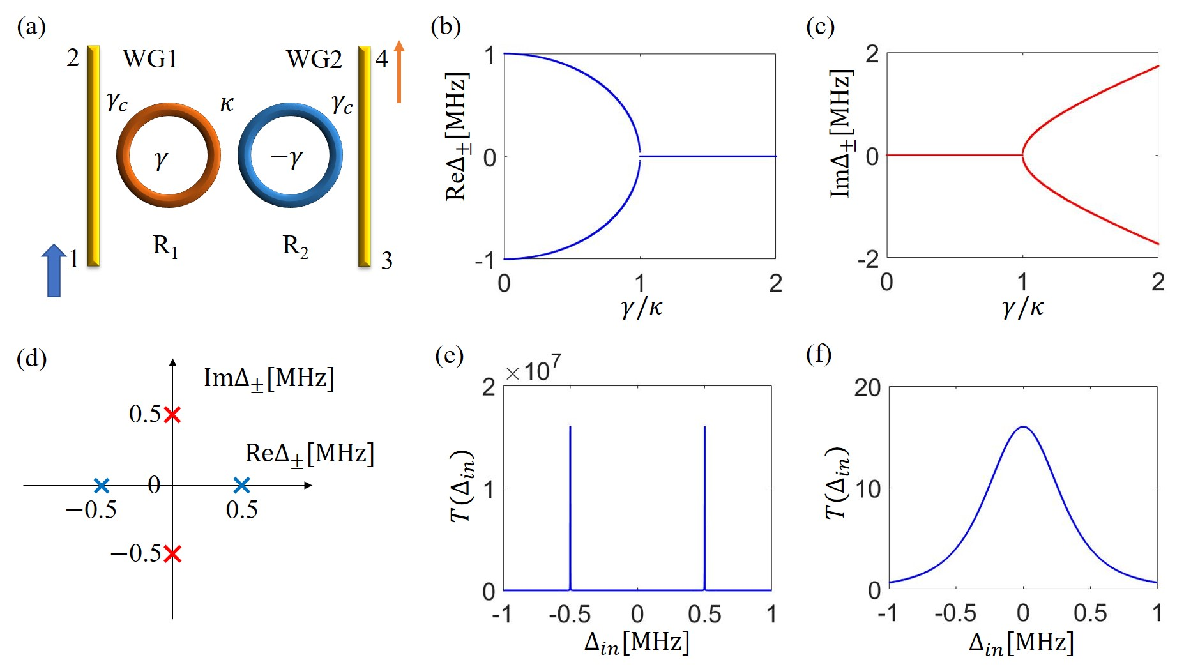}
		\caption{\footnotesize (a)The sketch of the two-mode $\mathcal{PT}$ symmetric system, wherein balanced gain(loss) is labeled by $\gamma$($-\gamma$), coupling between the two resonators is $\kappa$, and $\gamma_c$ denotes the couplings between the resonator ($R_1$ and $R_2$) and the waveguide (WG1 and WG2). (b) The real part of supermodes $\Delta_\pm$ of $\hat H_{2}$ as a function of $\gamma/\kappa$ with $\kappa>0$. (c) The imaginary part of $\Delta_\pm$ of $\hat H_{2}$ as a function of $\gamma/\kappa$ with $\kappa>0$. (d)The distribution of supermodes $\Delta_\pm$ on the complex plane where $\kappa=1$MHz. Blue points correspond to $\mathcal{PT}$ unbroken region ($\gamma=\sqrt{3}/2$MHz). Red points correspond to $\mathcal{PT}$ broken region ($\gamma=\sqrt{5}/2$MHz). (e)(f) Transmission spectrum against input frequency $\Delta_{in}$ in the $\mathcal{PT}$ unbroken ((e), $\gamma=\sqrt{3}/2$MHz) and broken ((f), $\gamma=\sqrt{5}/2$MHz) regions, wherein $\kappa=\gamma_c=1$MHz, $\epsilon=0.1$MHz.}\label{fig8}
	\end{figure}
\end{widetext}

\begin{figure}[htbp]
	\centering
	\includegraphics[width=0.44\textwidth]{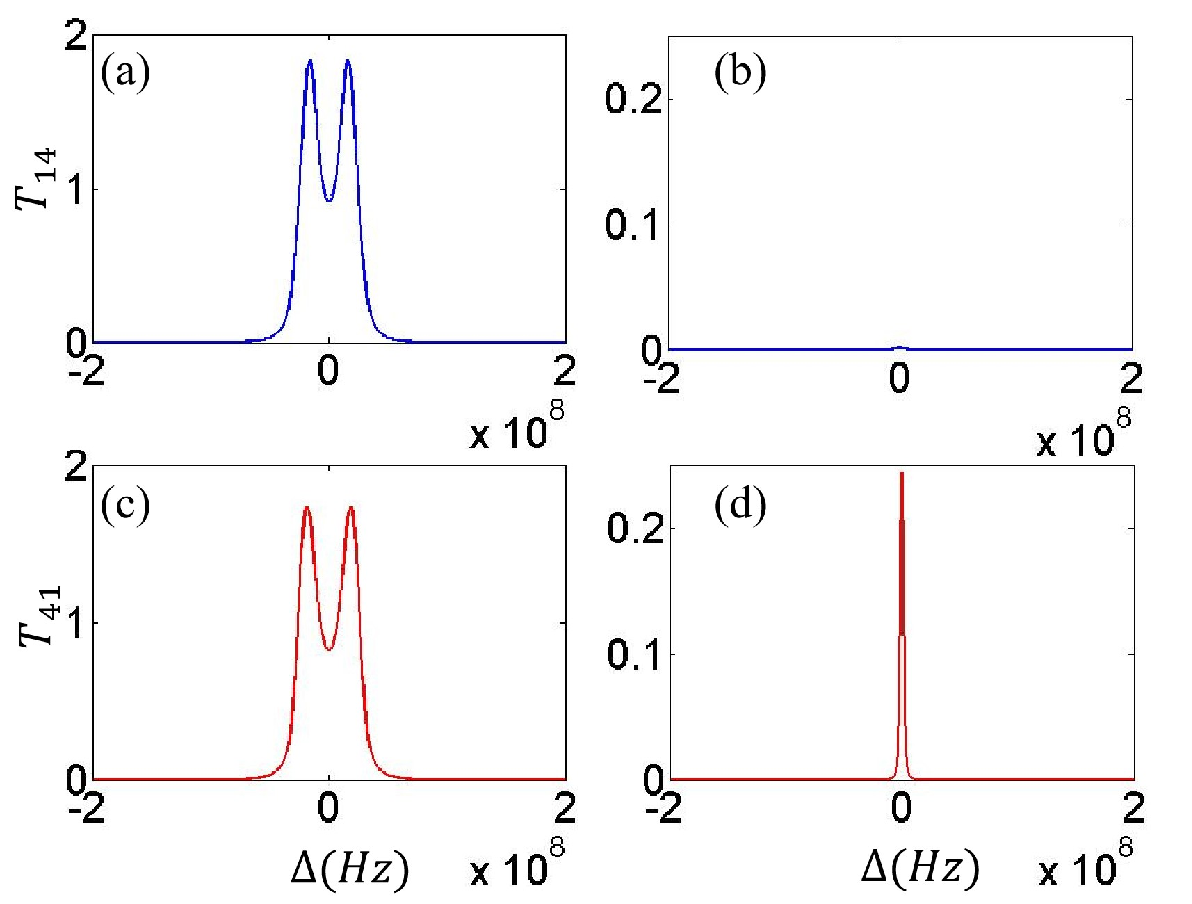}
	\caption{\footnotesize Transmission spectrum $T_{14}$ and $T_{14}$ against $\Delta_{\text{in}}$ wherein $\epsilon=176.05GH, B=0.01Hz$,
		(a)(c)$\gamma=3.766MH, \kappa=20MH$, (b)(e)$\gamma=3.766MH, \kappa=0.2MH$.}\label{fig9}
\end{figure}

One of the simplest examples is a two-mode array with balanced gain and loss, as shown in Fig.\ref{fig8}(a). This system can be easily built from the general structure as shown in Fig.\ref{fig2} and Eq.\eqref{aDyEq} by setting $N=2, \gamma_1=\gamma, \gamma_2=-\gamma$ and $\kappa_{1,2}=\kappa_{2,1}=\kappa$, where we have assumed $\gamma, \kappa>0$. Repeating the progress of Eqs.\eqref{aDyEq}-\eqref{H}, we can get the evolution matrix of the two-mode array:
\renewcommand\theequation{S1}
\begin{equation}\label{App1}
	\hat H_2=
	\left (
	\begin{aligned}
		\Delta_{in}&+i\gamma & & \kappa\\
		&\kappa & \Delta_{in}&-i\gamma\\
	\end{aligned}
	\right).
\end{equation}
According to Eq.\eqref{Trans}, the transmission spectrum can be given by
\renewcommand\theequation{S2}
\begin{equation}\label{App2}
	T(\Delta_{in})=\left|\frac{A_{out}}{A_{in}} \right|^2=\left|\frac{A_2}{\epsilon} \right|^2\gamma_c^2,
\end{equation}
where $A_2$ can be solved by $\alpha=\beta\hat H_2^{-1}$, wherein $\alpha=(A_1,A_2)^T$, $\beta=(\epsilon,0)^T$. In the complex plane, the distribution of supermodes as $\gamma/\kappa$ vary across the EP as shown in Fig.\ref{fig8}(d). When $\gamma/\kappa<1$ , the supermodes $\Delta_\pm$ lie on the real axis denoted by blue points. In this region, the transmission spectrum diverges at these points described in Fig.\ref{fig8}(e). When $\gamma/\kappa>1$, the values of $\Delta_\pm$ become purely imaginary as the red points in Fig.\ref{fig8}(d). Since the input detuning $\Delta_{in}$ is always a real number, the divergent peaks vanish in the transmission spectrum, and are replaced by a broad finite peak at $\Delta_{in}=0$ which is the projection of $\Delta_\pm$ on the real axis, as shown in Fig.\ref{fig8}(f).  In fact, according to $T(\Delta_{in})=|\kappa\gamma_c/(\Delta ^2_{in}+\gamma^2-\kappa^2)|^2$, it is easy to check that the spectrum can fits the square of a Breit-Wigner distribution around $\Delta_{in}=0$ when $\gamma^2-\kappa^2>0$. The peaks are called Breit-Wigner peaks~\cite{767}.

By considering the gain saturation effect, the linear equation can be reformulated into the following nonlinear form:
\renewcommand\theequation{S3}
\begin{equation}\label{App3}
	\begin{aligned}
		& \frac{d}{d t} A_1=i \Delta A_1+\gamma A_1-\kappa A_2-\frac{B}{2}\left|A_1\right|^2 A_1-\epsilon, \\
		& \frac{d}{d t} A_2=i \Delta A_2-\gamma A_2+\kappa A_1.
	\end{aligned}
\end{equation}
Comparing Eq.\eqref{ANDyEq} and Eq.\eqref{App3}, it can be seen that the nonlinear equation includes an additional cubic term. When the optical field inside the resonator is weak, that is $A_1$ being small, this term can be negligible due to the coefficient $B$ being very small. However, when the optical field increases, that is $A_1$ becoming larger, the cubic term increases dramatically, causing a substantial suppression of the field intensity, thus the field intensity cannot be linearly enhanced without limit.

Then we can calculate the transparency spectrum using this model, as shown in Fig.\ref{fig9}.  When $\gamma$ is small, which corresponds to the PT-symmetric unbroken phase depicted in Fig.\ref{fig8}(e), the original divergent peaks are constrained and transforms into non-divergent peaks, and exhibit reciprocity as shown in Fig.\ref{fig9}(a)(c). Conversely, when $\gamma$ is large, corresponding to the PT-symmetric broken phase shown in Fig.\ref{fig8}(f), the position of the original divergent peak remains unchanged, but the peak value of $T_{41}$ is significantly larger than that of $T_{14}$, demonstrating non-reciprocity, as shown in Fig.\ref{fig9}(b)(d). This agrees with the experimental results reported in ref.\cite{498}.

\end{document}